\documentclass[preprint]{aastex}

\usepackage{graphicx}
\usepackage{epstopdf}

\newcommand{\etal}{et~al.}

\newcommand{\rres}{r_{L}}

\newcommand{\Pone}{\citetalias{PorcoSOIScans06}}
\newcommand{\Ptwo}{\citetalias{soirings}}
\newcommand{\Eqn}[1]{Eq{#1}.}  
\newcommand{\Fig}[1]{Fig{#1}.}  
\defcitealias{PorcoSOIScans06}{Paper~I}
\defcitealias{soirings}{Paper~II}
\slugcomment{In Press at ApJL}

\shorttitle{Temporal Variability in Spiral Density Waves}
\shortauthors{Tiscareno \etal}

\begin{document}

\title{Unravelling Temporal Variability in Saturn's Spiral Density Waves: Results and Predictions}
\author{Matthew~S.~Tiscareno$^1$\footnote{Corresponding author:  \tt{matthewt@astro.cornell.edu}}, Philip~D.~Nicholson$^1$, Joseph~A.~Burns$^{1,2}$,\\ Matthew~M.~Hedman$^1$, Carolyn~C.~Porco$^3$}
\affil{$^1$ Department of Astronomy, Cornell University, Ithaca, NY 14853\\$^2$ Department of Theoretical and Applied Mechanics, Cornell University, Ithaca, NY 14853\\$^3$ CICLOPS, Space Science Institute, 4750 Walnut Street, Boulder, CO 80301}

\begin{abstract}
We describe a model that accounts for the complex morphology of spiral density waves raised in Saturn's rings by the co-orbital satellites, Janus and Epimetheus.  Our model may be corroborated by future Cassini observations of these time-variable wave patterns.  
\end{abstract}


\section{Introduction}

Owing to typical astronomical timescales, a galaxy's spiral arms are often considered as a fixed pattern.  So, too, for the numerous, tightly wound spiral waves detected in Saturn's rings.  In fact, both systems are dynamically active, with waves traveling away from resonant sites.  This is manifest only in Saturn's case, where a pair of moons -- Janus and Epimetheus -- occupy nearly identical orbits that are interchanged every 4 years, causing the resonance locations in the rings to skip back and forth by tens of km.  Since spiral density waves are initiated in Saturn's rings at locations where ring particle orbits are in a Lindblad resonance with a perturbing moon, the starting points of waves jump as well, allowing wave trains to interfere in complex ways \citep{PorcoSSR04}.  

High-resolution images of the rings were obtained by the Cassini spacecraft's Imaging Science Subsystem (ISS) on 2004~July~1 and on 2005~May~20/21.  The calibration and image processing of these data, resulting in a series of brightness scans with orbital radius, along with a catalog of important satellite gravitational resonances falling within the rings, are presented by \citet[hereafter \Pone]{PorcoSOIScans06}.  Further analysis of these data, employing techniques derived from the wavelet transform, is presented by \citet[hereafter \Ptwo]{soirings}.  

Examination of the Cassini images to date shows that density waves raised by the co-orbitals are both unusual and variable in their morphology.  In this paper, we describe a model that accounts for much of the observed structure.  We proceed to use this model to predict the future morphology of selected waves at the times and locations of planned Cassini observations, which we expect will test our predictions.  

\begin{figure}[!t]
\begin{center}
\includegraphics[width=8cm,keepaspectratio=true]{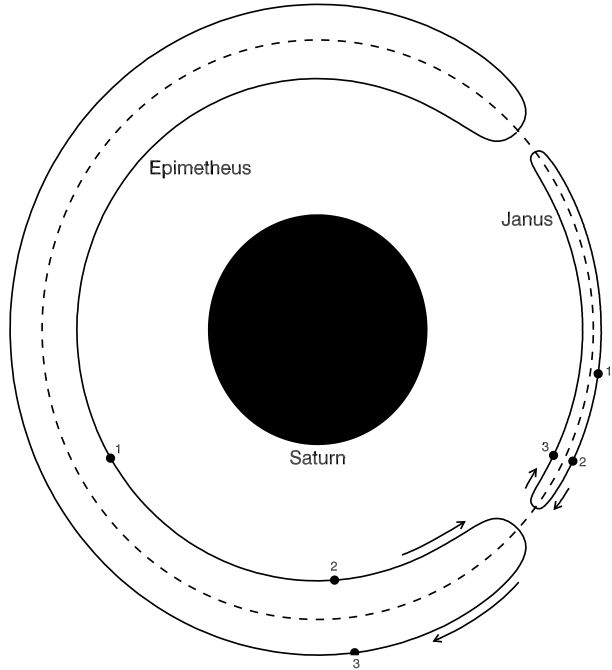}
\caption{The orbits of Janus and Epimetheus (with epicycles removed) in a frame rotating with their mass-averaged mean motion.  The radial separation of the librations is exaggerated by a factor of 500.  Numbered points mark the moons' positions on 1)~2004~July~1, 2)~2005~May~21, and 3)~2006~September~9.  Arrow lengths indicate motion accomplished in 100 days.  \label{JanEpiOrb}}
\end{center}
\end{figure}

\section{The Orbits of Janus and Epimetheus}

The orbits of Janus and Epimetheus about Saturn constitute a form of the three-body problem of celestial mechanics that is unique in the solar system \citep[see][]{DM81a,DM81b,Yoder83,Peale86}.  When viewed in a rotating frame of reference, whose angular velocity equals the mass-weighted average of the mean motions of the two moons, Epimetheus executes a modified ``horseshoe orbit'' encompassing Janus' slowly drifting $L_3$, $L_4$, and $L_5$ Lagrange points (\Fig{}~\ref{JanEpiOrb}).  However, since Epimetheus' mass is not negligible compared to Janus'  -- the mass ratio is 0.278 \citep{Spitale06} -- Janus executes its own libration about the average orbit.  

Because the orbits of the two moons are so similar ($\Delta a \sim 50$~km), they are commonly known as the ``co-orbital satellites'' (or, more briefly, the ``co-orbitals'').  Every 4.00 years, they execute their mutual closest approach and effectively ``trade'' orbits, the inner moon moving outward and vice versa.  The most recent reversal event occurred on 2006~January~21, at which time Janus became the inner satellite and Epimetheus the outer.  

\section{Spiral Density Waves \label{DW}}

Spiral density waves are raised in the rings at locations where ring-particle orbits are in a Lindblad resonance with a perturbing saturnian moon.  The density perturbations propagate outward from the resonance location.  As reviewed in \Ptwo{} \citep[see also][]{GT82,Shu84,NCP90,Rosen91a,Rosen91b,Spilker04}, five parameters characterize the idealized functional form of the perturbation:  1) the background surface density $\sigma_0$, which fixes the wavelength dispersion, 2) the resonance location $\rres$, specifying a linear translation of the wave, 3) the wave's initial phase $\phi_0$, 4) the damping parameter $\xi_D$, indicating a characteristic distance over which the wave propagates before damping away, and 5) the wave's amplitude $A_L$, which is proportional to the perturbing moon's mass.  

Considering only co-planar motions, the pattern speed of the resonant perturbation is described by positive integers $m$ and $k$+1; the first giving the number of spiral arms, and the second the order of the resonance (first-order being strongest).  A ($k$+1)th-order Lindblad resonance is generally labeled as $(m+k)$:$(m-1)$.  

At a given ring longitude $\lambda$, the initial phase of a particular density wave is 
\begin{equation}
\label{DWPhase}
\phi_0 = m \lambda - (m+k) \lambda_s + k \varpi_s, 
\end{equation}
\noindent where $\lambda_s$ and $\varpi_s$ are the perturbing moon's mean longitude and longitude of periapse, respectively. 

Although previous authors have usually analyzed spiral density waves in Saturn's rings as static phenomena, in fact they propagate with a finite group velocity \citep{Shu84}
\begin{equation}
v_g = \pi G \sigma_0 / \kappa , 
\end{equation}
\noindent where $\kappa$ is the radial (epicyclic) frequency of ring particle orbits, and $G$ is Newton's gravitational constant.  This is the speed at which information (e.g., effects of any change in the resonant forcing) propagates.  In the A~Ring, $v_g$ is on the order of 10-20~km/yr.  Since spiral density waves commonly propagate over many tens of km, and the forcing from the co-orbital satellites changes every 4 yr, discontinuities resulting from reversal events should be observable in density waves raised by the co-orbitals.  

\begin{figure}[!t]
\begin{center}
\includegraphics[width=7cm,keepaspectratio=true,viewport=0cm 0cm 8.5cm 8cm]{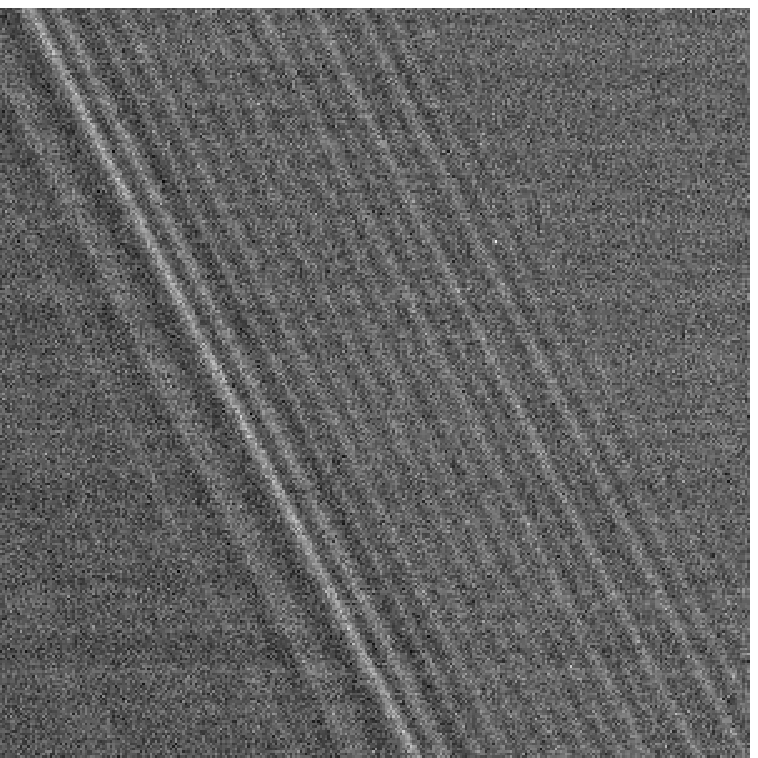}
\includegraphics[width=7cm,keepaspectratio=true,viewport=-1cm 0cm 11cm 11cm]{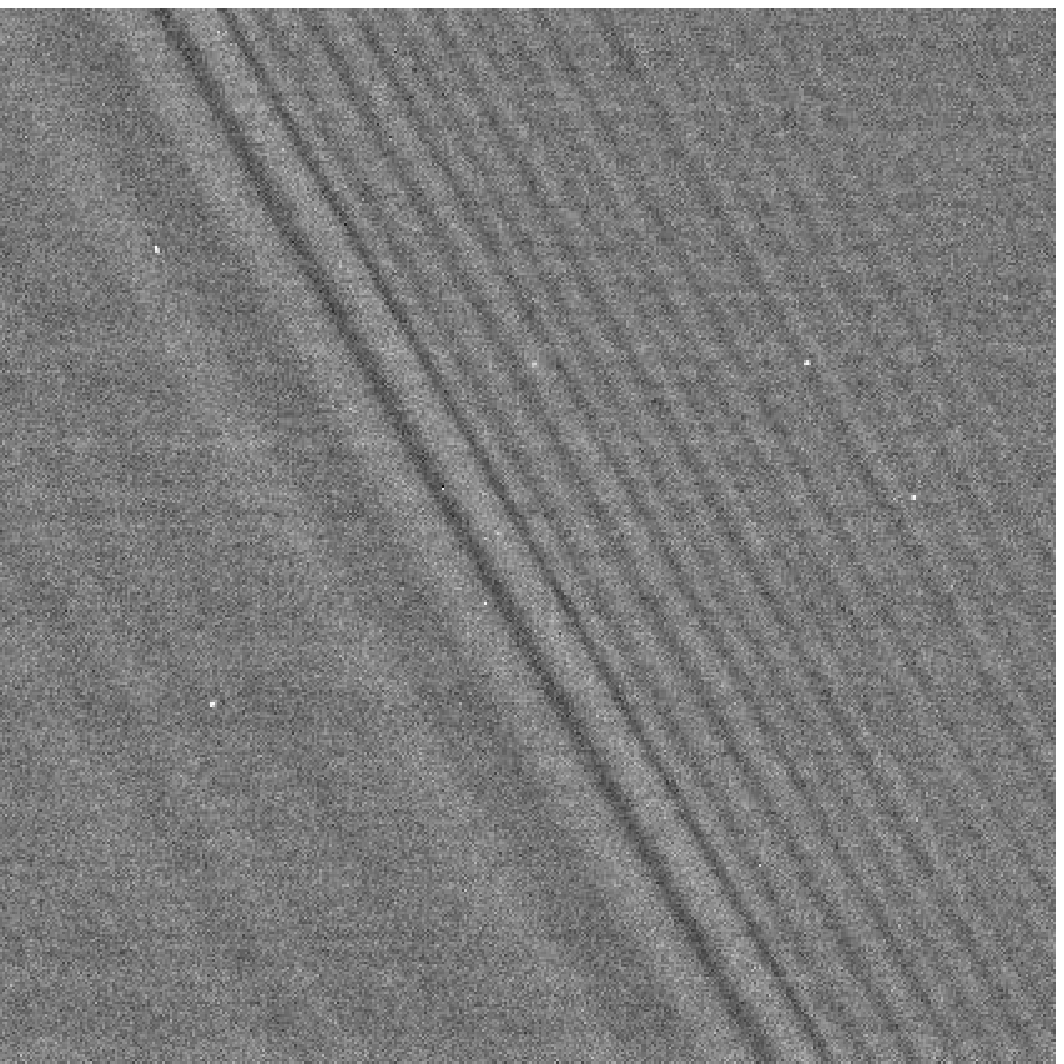}
\caption{Portions of Cassini images a) N1467345385 and b) N1467345916, showing spiral density waves raised by Janus and Epimetheus at the 7:5 and 9:7 resonances, respectively.  \label{JEWaveImages}}
\end{center}
\end{figure}

\section{Data}

In the radial scans of ring brightness taken from Cassini~ISS images (\Pone{}), spiral density waves due to the co-orbitals can be discerned at first-order, second-order, and third-order resonances.  

First-order waves (e.g., 2:1, 4:3, 5:4, 6:5) were seen by Voyager, and are some of the strongest waves in the ring system.  However, non-linear effects, which occur when the density perturbations are comparable to the background surface density, greatly complicate their analysis \citep{Shu85,BGT86,LB86}.  Not only does the wavelength dispersion deviate significantly from linear theory, but simple superposition of multiple wave segments is not valid.  

Second-order waves (e.g., 7:5, 9:7, 11:9, 13:11) were first clearly resolved in the present Cassini data set (\Fig{}~\ref{JEWaveImages}).  Since the density perturbations comprising these weaker waves are much smaller than the background surface density, they remain well-described by linear theory, and overlapping wave segments can be simply superposed.  Accordingly, second-order waves are best-suited for comparison with our simple model, and we will focus exclusively on them.  

Additionally, several third-order co-orbital waves can be discerned through wavelet analysis of the radial scans (\Ptwo{}); however, these are too weak for much detailed structure to be resolved.  

\section{Model}

Our model is based on a simple assumption:  When the co-orbital satellites are in a given configuration (we'll call the Janus-inward configuration ``JE'' and the Janus-outward configuration ``EJ''), spiral density waves propagate outward from the current Lindblad resonance locations at the group velocity $v_g$.  When a reversal occurs (e.g., the satellites changing configurations from JE to EJ), the JE resonance locations become inactive; however, the waves previously generated there continue to propagate outwards, resulting in a ``headless'' wave.  Meanwhile, new ``tailless'' spiral density waves begin to propagate outward at a speed $v_g$ from the EJ Lindblad resonance locations.

\begin{figure}[!t]
\begin{center}
\includegraphics[width=16cm,keepaspectratio=true]{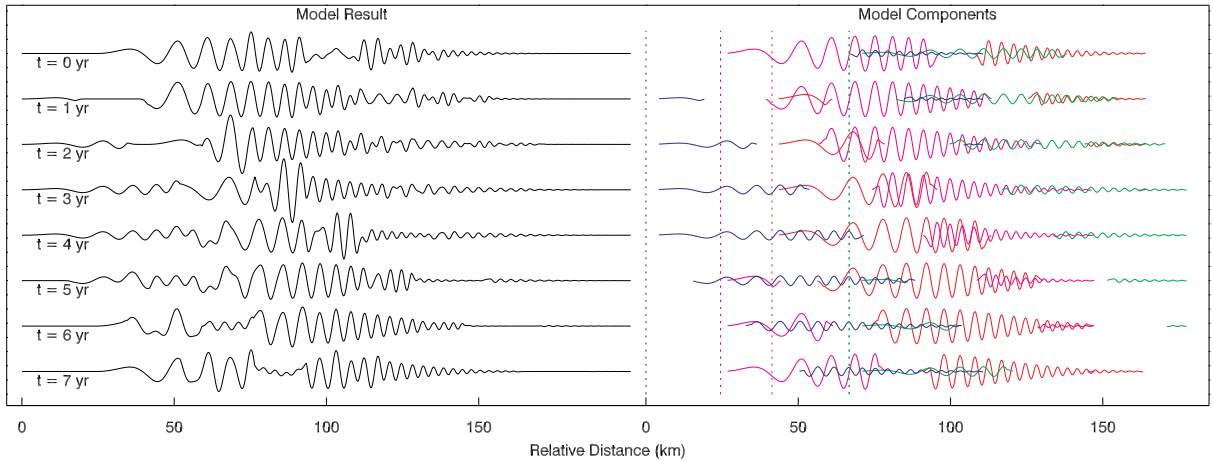}
\caption{An illustration of our model of co-orbital density waves, showing wave morphology at yearly intervals in the 8-yr libration cycle (reversal from JE to EJ occurs at $t=0$, and from EJ to JE at $t=4$~yr).  Parameters are based on the 9:7 density wave, but here $\phi_0 = 0$ for all wave segments.  In the right-hand column are wave segments created by Janus in its inner (purple) and outer (red) configurations, and Epimetheus in its inner (blue) and outer (green) configurations.  Only amplitudes larger than 1\% are shown.  Dotted lines show the resonance locations, with the same color coding.  The left-hand column shows the sum of the wave segments, which is the model prediction.  \label{JEModelDemo}}
\end{center}
\end{figure}

At any given time, the locations of ``headless'' and ``tailless'' wave segments can be easily calculated from $v_g$ and the elapsed interval since a reversal event.  We simply superpose these wave segments atop one another to arrive at a predicted wave morphology (\Fig{}~\ref{JEModelDemo}).  This time-domain approach to the problem is an alternative to the frequency-domain treatment of \citet{LGT85}.  

We set the relative amplitude to unity for Janus, and to the moons' mass ratio of 0.278 for Epimetheus.  The relevant resonance locations are easily derived from the moons' orbital parameters \citep[\Pone{}]{LC82}.  The absolute navigation of ISS images is not simple (see \Ptwo{}); nonetheless, all that is necessary for wave morphology is to know the separations among the resonance locations.  

The phase of each wave segment (\Eqn{}~\ref{DWPhase}) is determined by what the orbital parameters of the perturbing moon \textit{would have been} had no reversal taken place.  This is calculated using high-precision numerically-integrated orbits \citep{Spitale06}.  We make a linear fit to $\lambda_s$ and to $\varpi_s$ over a single inter-reversal time period, then extrapolate forward to the observation time.  

At the end of this process, the only remaining free parameters in our model are the background surface density $\sigma_0$ and the damping parameter $\xi_D$, which respectively control the wavelength dispersion and the propagation distance.  We manually adjust these parameters to find the optimum agreement of feature locations between model and data.  

For locations at which a wave segment begins or ends, an unrealistically sharp cutoff can occur.  We soften such discontinuities by using a half-gaussian, with a characteristic width of 1~km, to bring the perturbation back to zero.  

\begin{figure}[!t]
\begin{center}
\includegraphics[width=16cm,keepaspectratio=true]{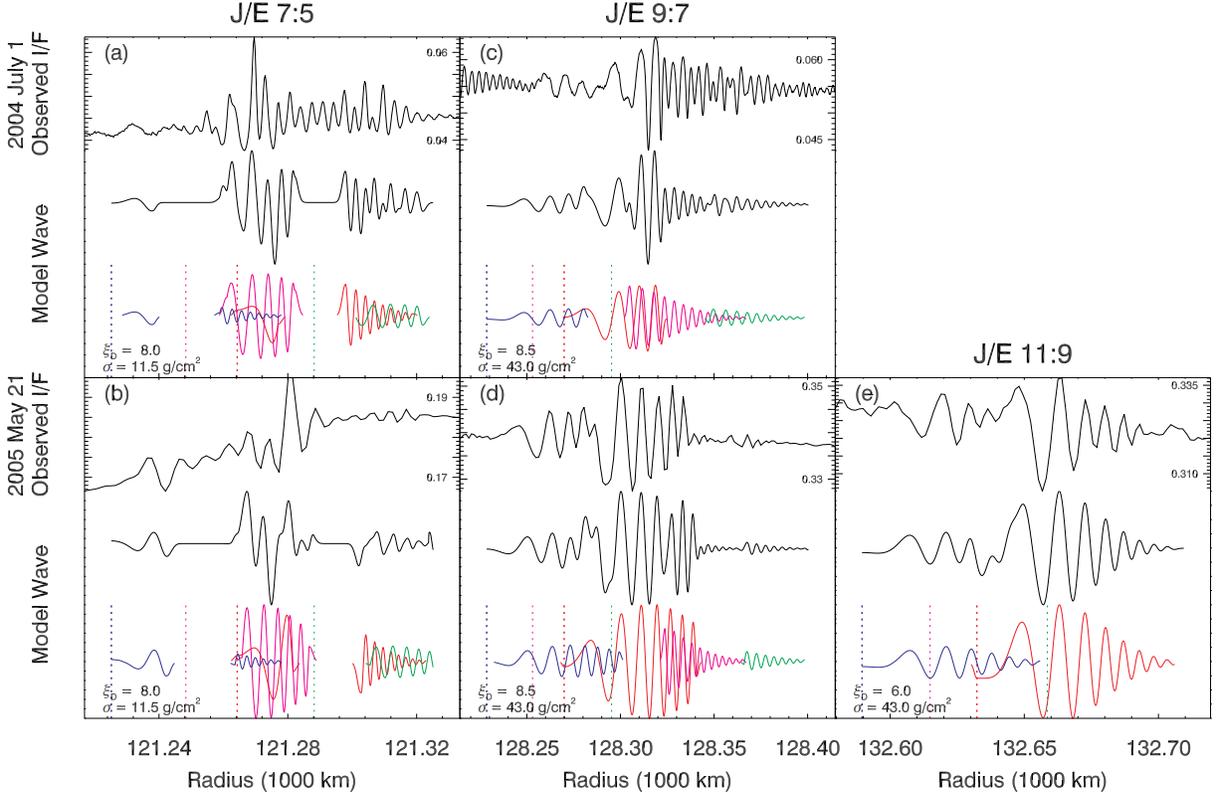}
\caption{Cassini data (\Pone{}) and model fits for selected Janus/Epimetheus density waves.  Resonance label, observation date, surface density $\sigma_0$, and damping parameter $\xi_D$ are given in each figure.  Radial traces were taken from the following images:  a) N1467345385, b) N1495327885, c) N1467345916, d) N1495326975, e) N1495326431.  Image resolutions are 250 m/pixel (a,c) and 1.4 km/pixel (b,d,e).  Model resolution has been degraded to correspond to image resolution.  \label{JEModelResults}}
\end{center}
\end{figure}

\section{Results}

Our model results are shown, along with corresponding Cassini image traces, in \Fig{}~\ref{JEModelResults}.  We find good qualitative agreement between the two, especially in the ``upstream'' (left-hand) regions.  There is also good agreement between model values of $\sigma_0$ and $\xi_D$ and values for nearby waves in the A Ring (9:7, 11:9) and Cassini Division (7:5) (\Ptwo).  We assume that image N1467345916 (\Fig{}~\ref{JEModelResults}c) is in the reverse-contrast regime (see \Ptwo{}).

The most glaring failure of the model occurs in regions where the predicted perturbation is zero (\Fig{s}~\ref{JEModelResults}c and~\ref{JEModelResults}d).  In such regions, the data instead show an oscillatory mode for which we cannot account.  It is possible that such oscillations could be raised in the several months during which the resonance locations are migrating from one configuration to the other, which we neglect in treating the reversals as instantaneous events.  Another potentially interesting explanation for such oscillations is that a leading spiral density wave may be propagating back towards the resonance location \citep{Shu84}.  This mode is allowed by the mathematical formalism, but has never been observed and has been considered impossible to excite; however, it is conceivable that a ``headless'' wave could send such a mode into an otherwise undisturbed region.  

\begin{figure}[!t]
\begin{center}
\includegraphics[width=16cm,keepaspectratio=true]{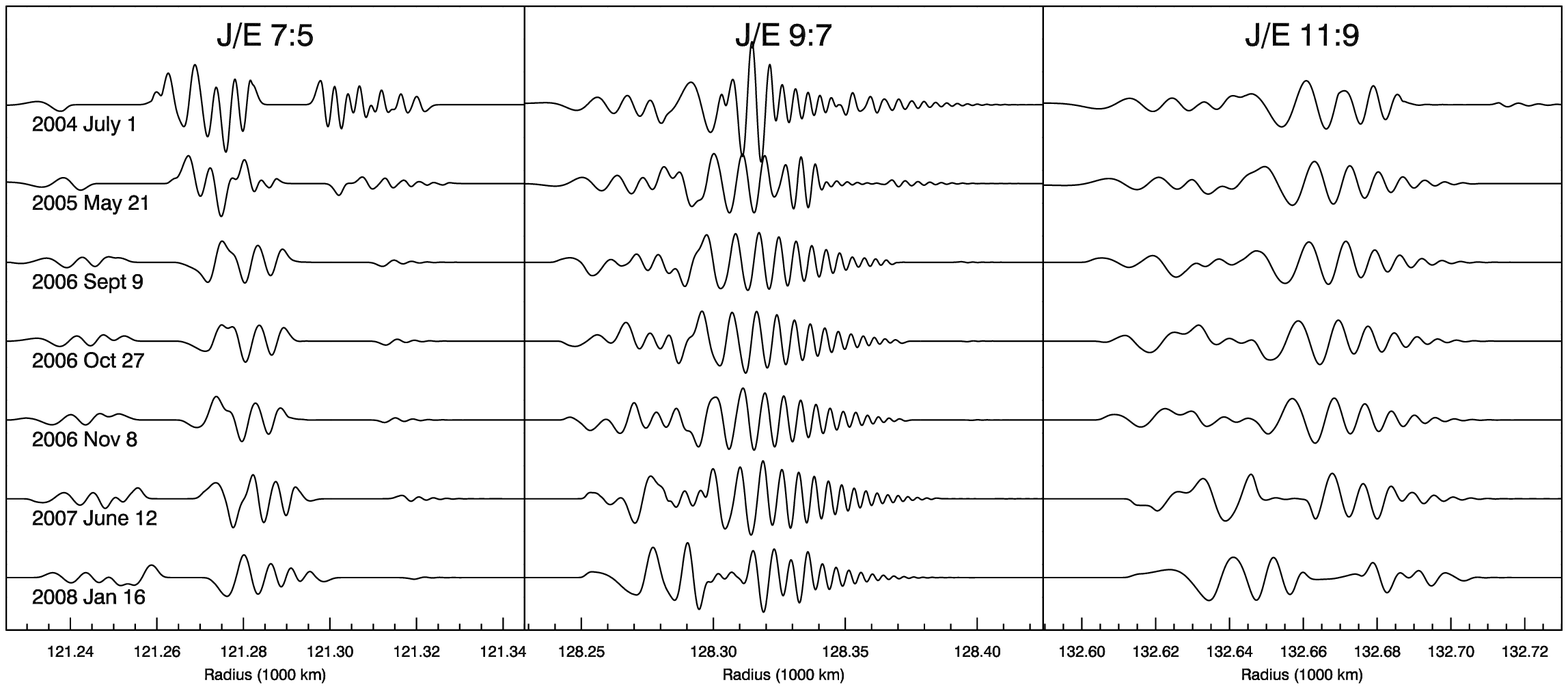}
\caption{Predicted morphology of Janus/Epimetheus density waves at times and locations of past and future Cassini~ISS observations.  Anticipated image resolutions are 700 m/pixel (2007~June~12) and 1.4 km/pixel (all others).  \label{JEModelPredict}}
\end{center}
\end{figure}

\section{Prediction}

Predictions of our model for the times and locations of Cassini~ISS observations designed to image the co-orbital waves \citep{PorcoSSR04} are shown in  \Fig{}~\ref{JEModelPredict}.  Owing to the spacecraft's low orbital inclination so far this year, the main rings have not been observed at high resolution since the last reversal event (from EJ to JE) on 2006~January~21, so the wave patterns in the new configuration have yet to be observed.  We predict that the innermost wavecycles (due to Epimetheus in the EJ configuration) will become headless and shift outwards, while a blank gap (possibly containing oscillations as in \Fig{s}~\ref{JEModelResults}c and~\ref{JEModelResults}d) will appear between the new inner Janus perturbation and the headless perturbation from the outer resonance location.  

The next reversal event, expected on 2010~January~21, will be observable only if the Cassini mission is extended beyond the current plan.  

\section{Conclusions}

Saturn's rings are a nearby and accessible natural laboratory \citep{BC06}, in which we can observe phenomena of broader interest, such as how a disk responds to variable forcing.  Our model is a first step towards a new understanding of the observed complex and time-variable waveforms.  We will revisit this topic in more detail, after the accuracy of our predictions (at least for data that will be obtained this year) becomes apparent.  

\acknowledgements{We thank R.~A.~Jacobson for providing the numerically-integrated satellite orbits.  We thank an anonymous reviewer for helping improve the manuscript.  We thank those involved in planning and executing the Cassini images, both at JPL and at CICLOPS.  We acknowledge funding by Cassini and by NASA PG\&G.}

\end{document}